\documentclass[preprintnumbers,amsmath,amssymb,aps,prl]{revtex4}
\usepackage{amsmath}
\usepackage[mathcal]{eucal}
\usepackage{latexsym}
\usepackage{amssymb}
\usepackage{color}
\newcommand*{\eps}{{\rlap{\lower2ex\hbox{$\,\,\tilde{}$}}{\epsilon_{ijk}}}}
\newcommand*{\EPS}{{\rlap{\lower2ex\hbox{$\,\,\tilde{}$}}{\epsilon_{i'j'k'}}}}
\newcommand*{\lmq}{{\rlap{\lower2ex\hbox{$\,\,\tilde{}$}}{\epsilon_{lmq}}}}
\newcommand*{\jmq}{{\rlap{\lower2ex\hbox{$\,\,\tilde{}$}}{\epsilon_{jmq}}}}
\newcommand*{\jql}{{\rlap{\lower2ex\hbox{$\,\,\tilde{}$}}{\epsilon_{jql}}}}
\newcommand*{\jlm}{{\rlap{\lower2ex\hbox{$\,\,\tilde{}$}}{\epsilon_{jlm}}}}
\newcommand*{\imq}{{\rlap{\lower2ex\hbox{$\,\,\tilde{}$}}{\epsilon_{imq}}}}
\newcommand*{\iql}{{\rlap{\lower2ex\hbox{$\,\,\tilde{}$}}{\epsilon_{iql}}}}
\newcommand*{\ilm}{{\rlap{\lower2ex\hbox{$\,\,\tilde{}$}}{\epsilon_{ilm}}}}
\newcommand*{\lmn}{{\rlap{\lower2ex\hbox{$\,\,\tilde{}$}}{\epsilon_{lmn}}}}
\newcommand*{\abc}{{\rlap{\lower2ex\hbox{$\,\,\tilde{}$}}{\epsilon_{abc}}}}
\newcommand*{\N}{{\rlap{\lower2ex\hbox{$\,\,\tilde{}$}}{N}}}
\newcommand{\tN}{{\rlap{\lower2ex\hbox{$\,\,\tilde{}$}}{N}}}
\newcommand*{\tM}{{\rlap{\lower2ex\hbox{$\,\,\tilde{}$}}{M}}}
\newcommand*{\imn}{{\rlap{\lower2ex\hbox{$\,\,\tilde{}$}}{\epsilon_{imn}}}}

\begin{document}
\title{Exact solutions of the Wheeler-DeWitt equation and the Yamabe construction}

\author{Eyo Eyo Ita III}\email{ita@usna.edu}
\address{Physics Department, US Naval Academy. Annapolis, Maryland}
\author{Chopin Soo}\email{cpsoo@mail.ncku.edu.tw}
\address{Department of Physics, National Cheng Kung University, Taiwan}
\input amssym.def
\input amssym.tex

\bigskip

\begin{abstract}
Exact solutions of the Wheeler-DeWitt equation of the full theory of  four dimensional
gravity of Lorentzian signature are obtained. They are characterized by Schr\"odinger wavefunctionals having support on 3-metrics of constant spatial scalar curvature,
and thus contain two full physical field degrees of freedom in accordance with the Yamabe construction. These solutions are moreover Gaussians of minimum uncertainty and they are naturally associated with a rigged Hilbert space.
In addition,  in the limit the regulator is removed, exact 3-dimensional diffeomorphism and local gauge invariance of the solutions are recovered.
\end{abstract}

\maketitle

\section{1. Introduction}

One of the major unsolved problems of theoretical physics is the consistent quantization of gravity; and solving the quantum constraints of General Relativity (GR) is among its most formidable challenges. In quantum theories without local gauge symmetries and constraints, eigenstates of the configuration space variable form a natural and easy basis, and every physical state can be expanded with respect to this basis. Dirac quantization of gravity requires physical states to be annihilated not just by kinematic generators of spatial diffeomorphisms, but also by the dynamical Wheeler-DeWitt operator\cite{DeWitt-Wheeler}. While classical analyses have revealed that there are 2 perturbative local degrees of freedom in Einstein's theory, construction of  the corresponding basis of explicit physical quantum states which also satisfy all the quantum constraints remains an intriguing and important task.  One of the greatest technical difficulties in solving the Wheeler-DeWitt constraint resides in balancing the potential and kinetic terms. A key breakthrough is the realization that the Yamabe theorem implies this can be very much simplified since the physical degrees of freedom in GR can be captured completely even if the spatial scalar curvature, which is the potential term in Einstein's theory, were to be restricted to be constant.

In this work we will provide a set of exact solutions of the full Wheeler-DeWitt equation, and construct a physical Hilbert space for four dimensional (GR) of Lorentzian signature which captures the essential 2 local gravitational degrees of freedom. We will carry out the canonical approach to quantization in the functional Schr\"odinger representation, and in the process shed light upon many of the unresolved issues.  While our initial motivation and guidance came from the affine group quantization formulation of GR \cite{SOOITACHOU}, these states can in fact be viewed as exact solutions of the conventional Wheeler-DeWitt equation expressed in terms of densitized triad-extrinsic curvature conjugate variables $({\tilde E}^{ia}, K_{ia})$ , and are not predicated upon only the affine group formulation.\par
\indent
In \cite{SOOITACHOU} a physical Hilbert space for gravity was proposed based upon the existence of unitary, irreducible representations of the affine group of transformations of the straight line $x\rightarrow{a}x+b$.  This result followed from the observation that the local Hamiltonian constraint of four dimensional gravity of Lorentzian signature can be written as an affine commutation relation involving Hermitian generators constructed from certain fundamental geometric objects
\begin{eqnarray}
\label{DELTA}
\widehat{H}(x)\vert\psi\rangle=\bigl([i\widehat{Q},\widehat{V}(x)]-\lambda\widehat{V}(x)\bigr)\vert\psi\rangle,
\end{eqnarray}
\noindent
where $\lambda=\frac{\hbar{G}\Lambda}{2}$ with $\Lambda$ being the cosmological constant.  Here, $iQ$ refers to the imaginary part of the Chern--Simons functional of the self-dual Ashtekar connection, and $V(x)$ is the local volume element of 3-space $\Sigma$, given by
\begin{eqnarray}
\label{VOLUME}
V(x)=\Bigl\vert\frac{1}{6}\widetilde{\epsilon}^{ijk}\epsilon_{abc}e^a_i(x)e^b_j(x)e^c_k(x)\Bigr\vert.
\end{eqnarray}
\noindent
In (\ref{VOLUME}) we have deviated from the standard practice in \cite{SOOITACHOU} by writing $V(x)$ not in terms of densitized triads $\widetilde{E}^i_a$, but rather in terms of the (undensitized) dreibein $e^a_i$, which is related by
\begin{eqnarray}
\label{TRIAD}
\widetilde{E}^i_a=\frac{1}{2}\widetilde{\epsilon}^{ijk}\epsilon_{abc}e^b_ie^c_j.
\end{eqnarray}
\noindent
The volume $V(x) =\sqrt{\vert\frac{1}{3!}\epsilon_{ijk}\epsilon_{abc}{\tilde E}^{ia}{\tilde E}^{jb}{\tilde E}^{kc}\vert}$ is polynomial in terms of the former variable and nonpolynomial in terms of the latter, a property which will be important for this paper\footnote{Index conventions for this work are that $a,b,c,\dots$ denote internal $SU(2)$ indices, while $i,j,k,\dots$ denote spatial indices in 3-space $\Sigma$.}.  The Chern--Simons functional of the Ashtekar self-dual connection  is given in differential form notation by the expansion
\begin{eqnarray}
\label{VOLUME1}
I_{CS}[A]=I_{CS}[\Gamma+iK]=Y+iQ=
I_{CS}[\Gamma]-\frac{1}{2}\hbox{tr}\int_{\Sigma}{k}\wedge{dk}
+i\hbox{tr}\Bigl(\int_{\Sigma}{F}\wedge{k}+\int_{\Sigma}{k}\wedge{k}\wedge{k}\Bigr),
\end{eqnarray}
\noindent
where we have defined the extrinsic curvature one form $k=\tau_aK^a_idx^i$, with $\tau_a$ being the $SU(2)$ generators and $\Gamma^a_i$ being the triad-compatible connection.   Since it is just the imaginary part rather than the full Chern--Simons functional, the Hermiticity of the genenerators in (\ref{DELTA}) is satisfied\cite{SOOITACHOU} . \par
\indent
 In this work our goal will be to provide an explicit representation of a physical Hilbert space $\textbf{H}_{Phys}$, as well as its explicit realization in terms of classical geometries encapsulating the essential gravitational degrees of freedom.  To harness the direct association to classical geometries, we will utilize the functional Schr\"odinger representation diagonalized on densitized triads with wavefunctionals $\psi[\widetilde{E}]=\langle\widetilde{E}\vert\psi\rangle$.  Upon carrying out the steps of the canonical quantization procedure we promote the initial value constraints to quantum operators which act on the state $\vert\psi\rangle$.  This entails a choice of a factor ordering, and for concreteness it is  affine algebra (\ref{DELTA}) which will form the basis for this input, but alternate factor-orderings of the Wheeler-DeWitt operator can also be easily incorporated into our scheme without altering the basic conclusions on the relation between the exact states and constant spatial scalar curvature manifolds. The scheme presented in this work can also be easily adapted from $(\tilde E^{ia}, K_{ia})$ to $( e_{ia}, {\tilde \pi}^{ia})$  or even metric $(q_{ij}, {\tilde \pi}^{ii})$ conjugate variables.\par

 The Hermiticity of the affine Lie algebraic generators in (\ref{DELTA}) means that the Hamiltonian constraint $\widehat{H}(x)$ in (\ref{DELTA}) is a self-adjoint operator.  Therefore we must use a self-adjoint operator ordering of the Wheeler--DeWitt version of the constraint in the functional Schr\"odinger representation. There are an infinite number of possible self-adjoint orderings for any composite operator, of which we will select one.  The imaginary part of the Chern--Simons functional can be read off directly from (\ref{VOLUME1}) and guided by aforementioned considerations, we are led to the natural ordering choice
\begin{eqnarray}
\label{DELTA1}
iQ=
i\int_{\Sigma}d^3y\Bigl(\frac{1}{2}\Bigl(\widetilde{R}^i_a[\Gamma]K^a_i+K^a_i\widetilde{R}^i_a[\Gamma]\Bigr)-\frac{1}{3!}\widetilde{\epsilon}^{ijk}\epsilon_{abc}K^a_iK^b_jK^c_k\Bigr).
\end{eqnarray}
\noindent
We will construct a set of functionals which are annihilated by the Wheeler--DeWitt equation which follows from the substitution of (\ref{DELTA1}) into (\ref{DELTA}), and interpret the ramifications of this choice with respect to solutions to the Einstein equations, as well as endow this space of solutions with an appropriate Hilbert space structure.\par
\indent
The organization of this paper is as follows.  In Section 2 we define the habitat space upon which the majority of the quantum field theory of this paper will be performed, namely, the functional Schwartz space.  This establishes once and for all, prior to performing any operations, that the operators exist on the appropriate spaces and that the operations to be carried out will be well-defined.  With this structure in place, in Sections 3 and 4 we perform an operator re-ordering of the Wheeler DeWitt equation into the desired self-adjoint form in preparation for transition into the functional Schrodinger representation.  In this representation, the wavefunctionals are diagonalized on the densitized triad.  In Section 5 we construct solutions for the initial value constraint operators of GR in the appropriate sense, and we provide a useful analogy to the harmonic oscillator to aid in the interpretation and the understanding of these results.  In Section 6 we provide the geometric meaning of the solutions with respect to solutions of the Einstein equations.  The wavefunctionals describe states peaked around configurations of constant three dimensional scalar curvature $R=6k$, which features two degrees of freedom per spatial point.  In Section 7 we analyse the Gaussian functional as a prototype for transition from the functional Schwartz space previousy defined to the physical, rigged Hilbert space.  The latter space consists of the exact solutions to the constraints, while the functional Schwartz space contains ``approximate solutions", approximate in a sense to be made precise by this paper.  The significance of the Gaussian is that it represents minimum uncertainty states with a well-known and well-defined Hilbert space structure, which as we will see, can in a sense be imported directly into the rigged Hilbert space formalism.  Section 8 is a brief discussion section, including directions of future research.

\section{2. Schwartz functional representation space}

To give meaning to the mathematical manipulations of this paper, we must choose a particular representation space for the states $\vert\psi\rangle$, upon which the action of all operators involved makes sense and is well-defined.  We will draw our motivation from the construction for rigged Hilbert spaces in ordinary quantum mechanics.  Here, one defines a Gelfand triple $\phi\subset{H}\subset\phi'$, where $H\in{L}^2(R),~dx$ are the square integrable functions.  The space $\phi$ is a dense subspace of $H$, upon which the action, expectation values and observables arising from unbounded operators are well-defined and preserve the space.\footnote{Only for bounded operators can the domain of the operators be chosen to be the whole Hilbert space.  For the unbounded case, the domain must be restricted to a subspace, usually dense in $H$, thereof.}  A particular example of $\phi$ is the Schwartz space $\boldsymbol{S}(R)$, namely the set of smooth functions of rapid decrease \cite{RIGGED}, \cite{RIGGED1}.  For all $f\in\boldsymbol{S}(R)$, we have that
\begin{eqnarray}
\label{TOPO}
\Vert{x}^m(d/dx)^nf\Vert<\infty,
\end{eqnarray}
\noindent
for $m$ and $n$ nonnegative integers.  The Schwartz space consists of the set of functions $f(x)$ which fall-off more rapidly than any polynomial in $x$ and its derivatives can blow up.  This space, in ordinary quantum mechanics, is preserved by all operators polynomial in $x$ and $p$, and will provide the motivation for its infinite dimensional analogue which we will define and utlize in this paper.  The space $\phi'$ is the topological dual of $\phi$ and includes the set of tempered distributions.  The space $\phi'$ can be seen as an enlargement of the Hilbert space $H$ to incorporate the so-called "generalized" functions.  Among this class include plane waves $e^{ipx}$ and Dirac delta functions $\delta(x-x')$, which the rigged Hilbert space formalism places upon a rigorous mathematical footing within the context of quantum mechanics. \cite{VON}, \cite{VON1}, \cite{VON2}.\par
\indent
We are now ready to define the habitat for the state $\vert\psi\rangle$ for this paper and their associated states of relevance.  Let $\Omega_x$ denote the set of densitized triad configurations $\widetilde{E}^i_a(x)$ at the spatial point $x\in\Sigma$, and define the set of infinitely functionally differentiable functionals of rapid decrease $\boldsymbol{S}(\Omega)$, via
\begin{eqnarray}
\label{TOPO1}
\boldsymbol{S}(\Omega)=\bigotimes_{x\in\Sigma}\boldsymbol{S}(\Omega_x),
\end{eqnarray}
\noindent
as the product of an infinite number of Schwartz spaces, one Schwartz space per spatial point $x$. To generalize (\ref{TOPO}) to the infinite dimensional spaces of field theory, we impose the following criterion which endows the functional Schwartz space $\boldsymbol{S}(\Omega)$ with a suitable topology
\begin{eqnarray}
\label{TOPO2}
\Vert\sum_{\sigma}P^{i_1i_2\dots{i}_mj_1j_2\dots{j}_n}_{a_1a_2\dots{a}_mb_1b_2\dots{b}_n}\prod_{k=1}^m\prod_{l=1}^n\Bigl(e^{a_k}_{\sigma(i_k)}(x)\frac{\delta}{\delta\widetilde{E}^{j_l}_{\sigma(b_l)}(x)}\Bigr)_{reg}\psi[\widetilde{E}]\Vert<\infty.
\end{eqnarray}
\noindent
The notation in (\ref{TOPO2}) is as follows.  We are summing over all permutations $\sigma$ not of indices but rather, of permutations of the positions of the operators taking into account all possible transposes.  Each spatial index $i_k(j_l)$ in a raised (lowered) position is paired with a corresponding internal
index $a_k(b_l)$ in the opposing lowered (raised) position such that the indices are linked together in all possible re-orderings.  The quantity $P$ is a tensor-valued object independent of the dynamical variables.\par
\indent
Note that (\ref{TOPO2}) is a local quantity defined at the point $x$.  In field theory the objects involved are usually operator-valued distributions whose local products are ill-defined.  In the field theoretical version of this paper, all triad-dependent quantities will act by multiplication on $\psi$ and any ultraviolet singularities would result from functional derivatives acting on triadic dependence at the same spatial point.  The abbreviation $reg$ signifies that these functional derivatives will be regularized by a point-splitting regulator $\tilde{f}_{\epsilon}(x,y)$, such that
\begin{eqnarray}
\label{REGOOL}
\int_{\Sigma}d^3y\tilde{f}_{\epsilon}(x,y)\varphi(y)=\varphi(x)
\end{eqnarray}
\noindent
for any field $\varphi$, where $\epsilon$ is a regularization parameter.  Each occurrence of a functional derivative acting at the same point within this regularization prescription will induce a factor of $\tilde{f}_{\epsilon}(0)$ in the coincidence limit $y\rightarrow{x}$.\par
\indent
The functional Schwartz space $\boldsymbol{S}(\Omega)$ can be though of for our purposes, loosely speaking, as a regularization space upon which all operators will be evaluated in their action on $\psi$.  To alleviate any potential concerns regarding regulator dependence of our final results, we will leave the specific form of the function $\tilde{f}_{\epsilon}(x,y)$ unspecified, other than to state the following property
\begin{eqnarray}
\label{DELTA7}
\hbox{lim}_{\epsilon\rightarrow{0}}\tilde{f}_{\epsilon}(x,y)=\delta^{(3)}(x,y).
\end{eqnarray}
\noindent
Note that $\delta^{(3)}(x,y)$ is a scalar of density weight one, which means that $\tilde{f}_{\epsilon}(x,y)\equiv{f}_{\epsilon}(x,\tilde{y})$ is really a scalar of density weight one with respect to $y$, the coordinate of integration in (\ref{REGOOL}).  So we must have for the coincidence limit that
\begin{eqnarray}
\label{DELTA8}
\hbox{lim}_{\epsilon\rightarrow{0},y\rightarrow{x}}\tilde{f}_{\epsilon}(x,y)=e(x)f_{\epsilon}(0).
\end{eqnarray}
\noindent
While $\tilde{f}_{\epsilon}(0)\rightarrow\infty$ as $\epsilon\rightarrow{0}$ there will be no infinities actually present in the space $\boldsymbol{S}(\Omega)$ for the following reasons: 1. First of all, due to the fall-off properties of $\psi\in\boldsymbol{S}(\Omega)$, $\psi$ will always approach zero faster than $f_{\epsilon}(0)$ and finite products of it can possibly blow up.  This is provided that the operators are polynomial in $e^a_i$ and in $\delta/\delta\widetilde{E}^i_a$, a condition which must be verified for all operators involved,\footnote{as we will see, each functional derivative $\delta/\delta\widetilde{E}^i_a$ with induce further factors of $e^a_i$ upon action either on the triads or on $\psi$.  This will be due in part to the density weight of the objects involved.} and is due to the fall-off properties of $\psi\in\boldsymbol{S}(\Omega)$.  For concreteness, one possible example of such a functional might be of the form $\psi[\widetilde{E}]=e^{-\alpha{f}_{\epsilon}(0)V}$, for some $\alpha>0$ and $V=\int_{\Sigma}d^3xV(x)$ with $V(x)$ given by (\ref{VOLUME}). 2. Secondly, in the limit of removal of the regulator we have $\epsilon=0$ and as we will see, the associated functionals $\psi$ corresponding to this limit will not exist on $\boldsymbol{S}(\Omega)$.  So the functional Schwartz space $\boldsymbol{S}(\Omega)$ is incomplete in the sense that the limit $\epsilon=0$ does not exist on this space.  So there will be no ultraviolet singularities on $\boldsymbol{S}(\Omega)$, and the highest power of occurrence a regulator possible in (\ref{TOPO2}) is of the form $(\tilde{f}_{\epsilon}(0))^n$, namely, it can be no higher than the number of functional derivatives present.\par
\indent
In this paper we will be implementing the Wheeler--DeWitt equation, schematically of the form $e(R/2+\Lambda)+KKe+\dots$, where the dots denote re-ordering terms, which consist of unbounded operators acting on $\vert\psi\rangle$.  The condition (\ref{TOPO2}) is clearly satisfied for the momentum squared term, which is quadratic in $K^a_i$ (and hence quadratic in functional derivatives) and linear in the triad $e^a_i$.  The triadic curvature is given by
\begin{eqnarray}
\label{CURVATURE}
\widetilde{R}^i_a[\Gamma]=\widetilde{\epsilon}^{ijk}\partial_j\Gamma^a_i+\frac{1}{2}\widetilde{\epsilon}^{ijk}\epsilon_{abc}\Gamma^b_j\Gamma^c_k,
\end{eqnarray}
\noindent
where the triad-compatible spin connection $\Gamma^a_i$, which depends explicitly on the dreibein $e^a_i$, is given by
\begin{eqnarray}
\label{CURVATURE1}
\Gamma^b_m=\frac{1}{e}\Bigl(\widetilde{\epsilon}^{ijk}e^b_ie^a_m\partial_je^a_k-\frac{1}{2}e^b_m\widetilde{\epsilon}^{njk}e^a_n\partial_je^a_k\Bigr).
\end{eqnarray}
\noindent
One sees in (\ref{CURVATURE}) that the curvature term is polynomial in $\Gamma^a_i$, which from (\ref{CURVATURE1}) is in turn polynomial in $e^a_i$ and its (spatial) derivatives.  Since $\widetilde{R}^i_a[\Gamma]$ acts multiplicatively on $\psi[\widetilde{E}]$, then this term also satisfies (\ref{TOPO2}), thus preserving the functional Schwartz space $\boldsymbol{S}(\Omega)$.  We will see in the course of the paper that all objects brought down from $\psi$ via functional derivatives in the Hamiltonian constraint will be polynomial in $e^a_i$.  So the kinetic term of the Hamiltonian constraint will respect (\ref{TOPO2}), thus preserving $\boldsymbol{S}(\Omega)$.\footnote{The analogue for ordinary quantum mechanics of a failure to meet this requirement would be operatio resulting in $m<0$, as well as noninteger exponents. }\par
\indent
We have presented arguments that the Hamiltonian constraint preserves the habitat representation space $\psi[\widetilde{E}]$, with respect to multiplication by (undensitized triads) and functional differentiation with respect to densitized triads $\widetilde{E}^i_a$.  So all formal operations in what follows in this paper will be mathematically rigorous and well-defined due to (\ref{TOPO2}).  So the next step is to expand out the affine algebra (\ref{DELTA}).  Prior to proceeding it is important to mention that the Hilbert space $\boldsymbol{H}_{Kin}$, the field theoretical analogue corresponding to the space of which the functional Schwartz
space $\boldsymbol{S}(\Omega)$ is a dense subspace, is
\begin{eqnarray}
\label{SUBSPACE}
\boldsymbol{H}_{Kin}=\otimes_{x\in\Sigma}{L}_2(\Omega_x,\delta\mu_x(\widetilde{E}),
\end{eqnarray}
\noindent
the square integrable functionals of the densitized triad $\widetilde{E}^i_a(x)$ with respect to the measure $\delta\mu_x(\widetilde{E})$.  We have an infinite product of $L_2$ spaces, one $L_2$ space per spatial point $x$.  The space $\boldsymbol{H}_{Kin}$ serves in the role of a kinematical Hilbert space, namely a Hilbert space at the level prior to solving any constraints.  The space $\boldsymbol{H}_{Kin}$ will not be actually utilized in this paper, and we have mentioned it for the sake of completeness.\footnote{As a prototypical example we invoke the infinite dimensional abstract Wiener spaces in field theory.  The functional Schwartz space, restricted to the case of Wiener measure, takes on the role of the Cameron--Martin space in measure theory.  The Camerion--Martin space is densely embedded in a Hilbert space, here serving as the kinematic Hilbert space $\textbf{H}_{Kin}$, via the inclusion map $\imath(\boldsymbol{S}(\Omega))\subset\textbf{H}_{Kin}$.  Functional integration is performed on this space with respect to an infinite dimensional Wiener measure, which is essentially a Gaussian measure on an infinite product of continuous paths.}\par
\indent

\section{3. Self-adjoint re-ordering of the Wheeler--DeWitt equation}

Having defined the habitat space which will be needed and having established its desired properties, we are now ready to proceed with writing down the self-adjoint ordering for the Wheeler--DeWitt equation desired.
By application of the quantum form of the Leibniz rule, the commutator term of  Hamiltonian constraint (\ref{DELTA}) is given by
\begin{eqnarray}
\label{DELTA2}
\frac{i}{2}\int_{\Sigma}d^3y\Bigl(\widetilde{R}^i_a[\Gamma(y)][\widehat{K}^a_i(y),\widehat{V}(x)]+[\widehat{K}^a_i(y),\widehat{V}(x)]\widetilde{R}^i_a[\Gamma]\Bigr)
-\frac{i}{3!}\int_{\Sigma}d^3y\widetilde{\epsilon}^{ijk}\epsilon_{abc}\Bigl(\widehat{K}^a_i(y)\widehat{K}^b_j(y)[\widehat{K}^c_k(y),\widehat{V}(x)]\nonumber\\
+\widehat{K}^a_i(y)[\widehat{K}^b_j(y),\widehat{V}(x)]\widehat{K}^c_k(y)+[\widehat{K}^a_i(y),\widehat{V}(x)]\widehat{K}^b_j(y)\widehat{K}^c_k(y)\Bigr).
\end{eqnarray}
\noindent
Note that in light of (\ref{CURVATURE}) and (\ref{CURVATURE1}), it is possible to arrive at a more complicated self-adjoint ordering than (\ref{DELTA1}) for the curvature term by averaging over all possible transposes of $K^a_i$ with respect to the individual triads.  Since commutation with $V(x)$ replaces all occurrences of $K^a_i$ with $e^a_i$ which commutes with the remaining terms, then only the curvature $eR$ can result, regardless of the initial ordering.
Utilizing the following commutation relation
\begin{eqnarray}
\label{DELTA3}
[\widehat{K}^a_i(x),\widehat{V}(y)]=\frac{i}{2}(\hbar{G})e^a_i\delta^{(3)}(x,y),
\end{eqnarray}
\noindent
then (\ref{DELTA2}) is given by\footnote{We have omitted the hats from the triads $e^a_i$, since these will act by multiplication in the funtional Schr/"odinger representation diaganolized on densitized triads.}
\begin{eqnarray}
\label{DELTA4}
-\frac{1}{2}(\hbar{G})e^a_i\widetilde{R}^i_a[\Gamma]
+\frac{1}{12}(\hbar{G})\widetilde{\epsilon}^{ijk}\epsilon_{abc}\Bigl(\widehat{K}^a_i\widehat{K}^b_je^c_k
+\widehat{K}^a_ie^b_j\widehat{K}^c_k+e^a_i\widehat{K}^b_j\widehat{K}^c_k\Bigr).
\end{eqnarray}
\noindent
Note that (\ref{DELTA4}) is a Weyl-ordered form of the density weight one constraint, which is polynomial when expressed in terms of the (undensitized) triad $e^a_i$.  However, in terms of the Ashtekar densitized inverse triad $\widetilde{E}^i_a$, a density of weight one, we would have a nonpolynomial Hamiltonian constraint.  Nonpolynomiality of the density weight one form of the Hamiltonian constraint has been argued, for instance by Thiemann\cite{THIEMANN}, to be the price that has to be paid in exchange for its desirable self-regulating properties, with the converse holding true for the density weight two form.  Our particular choice of variables is designed to capitalize upon the optimal features of both worlds, and as well respects the requirement (\ref{TOPO2}).\par
\indent
Let us focus first on the momentum squared terms of (\ref{DELTA4}), given by
\begin{eqnarray}
\label{DELTA5}
\frac{1}{12}(\hbar{G})\widetilde{\epsilon}^{ijk}\epsilon_{abc}\Bigl(\widehat{K}^a_i\widehat{K}^b_je^c_k
+\widehat{K}^a_ie^b_j\widehat{K}^c_k+e^a_i\widehat{K}^b_j\widehat{K}^c_k\Bigr).
\end{eqnarray}
\noindent
Using the relation
\begin{eqnarray}
\label{DELTA6}
[\widehat{K}^a_i(y),e^b_j(x)]=i(\hbar{G})\bigl(\frac{1}{2}e^a_ie^b_j-e^b_ie^a_j\bigr)\frac{1}{e}\delta^{(3)}(x,y),
\end{eqnarray}
\noindent
we will re-order the terms of (\ref{DELTA5}) into the self-adjoint $KeK$ form.  Due to the nonlinearity of the Hamiltonian constraint in momenta, with products of operator-valued distributions at the same spatial point, we will need to put in place some sort of regularization prescription.\footnote{It is often said that the introduction of a regulator by default breaks diffeomorphism invariance in quantum gravity, where remnants of the regulator persist even after it has been removed.  The results of this paper will shed light upon the precise mechanisms underlying this this ailment as well as its cure.}  To this end let us introduce a point-splitting regulator $\tilde{f}_{\epsilon}(x,y)$, satisfying (\ref{DELTA7}) and (\ref{DELTA8})
\noindent
picking up an explicit factor of $e(x)=V(x)$.  Recall that $f$ without a tilde is a c-number of density weight zero.\par
\indent
The middle term of (\ref{DELTA5}) is already of the required self-adjoint form, so we need only reorder the first and the third terms.  The first term can be written, after a relabelling of indices and using the commutation relations as
\begin{eqnarray}
\label{DELTA9}
\widehat{p}(x)=\frac{1}{12}(\hbar{G})\widetilde{\epsilon}^{ijk}\epsilon_{abc}\widehat{K}^a_i\widehat{K}^c_ke^b_j
=\frac{1}{12}(\hbar{G})\widetilde{\epsilon}^{ijk}\epsilon_{abc}\bigl(\widehat{K}^a_ie^b_j\widehat{K}^c_k+\widehat{K}^a_i[\widehat{K}^c_k,e^b_j]\bigr).
\end{eqnarray}
\noindent
To avoid singularities due to the coincidence limit of (\ref{DELTA6}), we will perform a point-splitting regularization of the commutator to
\begin{eqnarray}
\label{DELTA10}
\frac{1}{12}(\hbar{G})\widetilde{\epsilon}^{ijk}\epsilon_{abc}\widehat{K}^a_i(x)\int_{\Sigma}d^3y\tilde{f}_{\epsilon}(x,y)[\widehat{K}^c_k(y),e^b_j(x)]\nonumber\\
=\frac{1}{12}(\hbar{G})\widetilde{\epsilon}^{ijk}\epsilon_{abc}\widehat{K}^a_i(x)\int_{\Sigma}d^3y\tilde{f}_{\epsilon}(x,y)
i(\hbar{G})\Bigl(\frac{1}{2}e^b_je^c_k-e^c_je^b_k\Bigr)\frac{1}{e}\delta^{(3)}(x,y)\nonumber\\
=i\frac{(\hbar{G})^2}{12}\widetilde{\epsilon}^{ijk}\epsilon_{abc}\widehat{K}^a_i\Bigl(\frac{3}{2}e^b_je^c_k\Bigr)f_{\epsilon}(0)
=\frac{i(\hbar{G})^2f_{\epsilon}(0)}{4}\widehat{K}^a_i\widetilde{E}^i_a
\end{eqnarray}
\noindent
where we have used (\ref{DELTA6}) and $f_{\epsilon}(0)=\tilde{f}_{\epsilon}(0)/e(x)$.  So the regularized form of (\ref{DELTA9}) is given by
\begin{eqnarray}
\label{DELTA11}
\widehat{p}_{\epsilon}(x)\vert\psi\rangle=
\Bigl(\frac{1}{12}\widetilde{\epsilon}^{ijk}\epsilon_{abc}\widetilde{K}^a_ie^b_j\widetilde{K}^c_k+
\frac{i(\hbar{G})^2f_{\epsilon}(0)}{4}\widehat{K}^a_i\widetilde{E}^i_a\Bigr)\vert\psi\rangle,
\end{eqnarray}
\noindent
with the densitized triad $\widetilde{E}^i_a$ appearing to the right.  Repeating the analogous previous steps, it is not difficult to see that the corresponding calculation for the third term of (\ref{DELTA5}) will result in
\begin{eqnarray}
\label{DELTA12}
\widehat{q}_{\epsilon}(x)\vert\psi\rangle=
\Bigl(\frac{1}{4}(\hbar{G})^2\widetilde{\epsilon}^{ijk}\epsilon_{abc}\widetilde{K}^a_ie^b_j\widetilde{K}^c_k-
\frac{i(\hbar{G})^2f_{\epsilon}(0)}{4}\widetilde{E}^i_a\widehat{K}^a_i\Bigr)\vert\psi\rangle.
\end{eqnarray}
\noindent
Note the minus sign in (\ref{DELTA12}) due to reversal of the terms in the commutator, as well as the fact that the momentum term in the second term now appears on the right.  So we can write down a regularized form of (\ref{DELTA5}), given by its middle $KeK$ term combined with (\ref{DELTA11}) and (\ref{DELTA12}), which yields
\begin{eqnarray}
\label{DELTA13}
\widehat{c}_{\epsilon}(x)\vert\psi\rangle
=\frac{1}{4}(\hbar{G})\Bigl(\widetilde{\epsilon}^{ijk}\epsilon_{abc}\widetilde{K}^a_ie^b_j\widetilde{K}^c_k
+i(\hbar{G}f_{\epsilon}(0))(\widehat{K}^a_i\widetilde{E}^i_a-\widetilde{E}^i_a\widehat{K}^a_i\bigr)\Bigr)\vert\psi\rangle.
\end{eqnarray}
\noindent
We have acquired another commutator of terms evaluated at the same point $x$, this time involving the densitized triad $\widetilde{E}^i_a$.  Since this is polynomial in the triad via (\ref{TRIAD}), then this commutator evaluated explicitly will also respect (\ref{TOPO2}).  So we can direclty make use of the commutation relation
\begin{eqnarray}
\label{DELTA14}
[\widehat{K}^a_i(x),\widetilde{E}^j_b(y)]=i(\hbar{G})\delta^a_b\delta^j_i\delta^{(3)}(x,y).
\end{eqnarray}
\noindent
It is not difficult to see, upon using the previous regularization prescription, that in the coincidence limit we will get
$[\widehat{K}^a_i,\widetilde{E}^i_a]\rightarrow{9}i(\hbar{G}f_{\epsilon}(0))$.  Putting this result into (\ref{DELTA13}) and using the relation
\begin{eqnarray}
\label{DELTA15}
\widetilde{R}^i_a[\Gamma]e^a_i=\frac{eR}{2},
\end{eqnarray}
\noindent
where $R$ is the three dimensional spatial Ricci scalar of the 3-metric $h_{ij}=e^a_ie^a_j$, then incorporating a cosmological constant term into (\ref{DELTA5}) and cancelling off a factor of $\frac{1}{2}(\hbar{G})$, we can write down the following self-adjoint, regularized Hamiltonian constraint descended from the affine algebra (\ref{DELTA}) for the ordering (\ref{DELTA1}), given by
\begin{eqnarray}
\label{DELTA16}
\widehat{H}_{\epsilon}(x)\vert\psi\rangle
=\biggl[e\Bigl(\frac{R}{2}+\Lambda+\frac{9}{2}(\hbar{G}f_{\epsilon}(0))^2\Bigr)-\frac{1}{2}\widetilde{\epsilon}^{ijk}\epsilon_{abc}\widehat{K}^a_ie^b_k\widehat{K}^c_k\biggr]\vert\psi\rangle.
\end{eqnarray}
\noindent
Starting from the affine algebra (\ref{DELTA}) and upon reordering operators in preparation for the densitized triad representation, we have obtained a regulator-dependent contribution to the cosmological constant which would blow up if the the regulator were to be naively removed.  This will need to be dealt with, and in order to identically satisfy the Hamiltonian constraint $\widehat{H}(x)$, we need to find a functional which, in an appropriately rigorous sense, is annihilated by the constraint after the regulator has been removed.

\section{4. Quantization in the functional Schr\"odinger representation}

We need to find a set of gauge-invariant, diffeomorphism-invariant functionals which, upon removal of the regulator, is exactly annihilated by the Wheeler--DeWitt equation (\ref{DELTA16}) which follows from (\ref{DELTA}) and (\ref{DELTA1}).  Denote by $\boldsymbol{Z}_{e_R}(\Omega)$, the set of rapidly descreasing functionals in $\psi[\widetilde{E}]\in\boldsymbol{S}(\Omega)$ which are  peaked on particular triad configurations $e_R$.  We will show that wavefunctionals $\psi[\widetilde{E}]\in\boldsymbol{Z}_{e_R}(\Omega)$ satisfy this condition.  We will see that the required wavefunctional will have support on triads having a numerically constant three dimensional spatial curvature scalar $R$.  Prior to displaying an explicit example of the required functional, we will first develop its desired properties through physical arguments and motivations.  Let us initially assume a wavefunctional of the form
\begin{eqnarray}
\label{ASSUME}
\psi[\widetilde{E}]=e^{S[\widetilde{E}]},
\end{eqnarray}
\noindent
leaving off any normalization factors for the time being, where $S$ is a functional to be determined.  We require that $\psi\in\boldsymbol{Z}_{e_R}(\Omega)\subset\boldsymbol{S}(\Omega)$ is an element of the functional Schwartz space, as defined in (\ref{TOPO1}).  We must now evaluate the Hamiltonian constraint (\ref{DELTA16}) acting on $\psi[\widetilde{E}]=\langle\widetilde{E}\vert\psi\rangle$.  We will need the equal-time commutation relations
\begin{eqnarray}
\label{COMMU}
[\widehat{\widetilde{E}^i_a}(x),\widehat{\widetilde{E}^j_b}(y)]=[\widehat{K}^a_i(x),\widehat{K}^b_j(y)]=0;~~
[\widehat{K}^a_i(x),\widehat{\widetilde{E}^j_b}(y)]=i(\hbar{G})\delta^a_b\delta^j_i\delta^{(3)}(x,y).
\end{eqnarray}
\noindent
We now make the third main input for our results aside from the affine algebra and the functional Schwartz space, namely the particular functional Schr\"odinger representation to be utilized, from the infinite range of possibilities inherent in (\ref{COMMU}).\footnote{That is, we can add an arbitrary functional of $\widetilde{E}^i_a$ to the mometum $K^a_i$ and still satisfy (\ref{COMMU}).}  So densitized triads will act via multiplication and momenta via functional differentiation
\begin{eqnarray}
\label{DELTA22}
\widehat{\widetilde{E}^i_a(x)}\psi[\widetilde{E}]=\widetilde{E}^i_a(x)\psi[\widetilde{E}];~~
\widehat{K}^a_i(x)\psi[\widetilde{E}]=i(\hbar{G})\frac{\delta\psi[\widetilde{E}]}{\delta\widetilde{E}^i_a(x)}.
\end{eqnarray}
\noindent
For the wavefunctional we have chosen the form (\ref{ASSUME}) specifically for its convenience in the evaluation of the semiclassical limit, which will be important for our analysis.  From (\ref{COMMU}), one can readily read off the extrinsic curvature as
\begin{eqnarray}
\label{COMMU1}
\widehat{K}^a_i(x)\psi[\widetilde{E}]=i(\hbar{G})\Bigl(\frac{\delta{S}}{\delta\widetilde{E}^i_a(x)}\Bigr)\psi[\widetilde{E}].
\end{eqnarray}
\noindent
The result of acting with the triadic extrinsic curvature operator yields its Hamilton--Jacobi functional value times $\psi$.  Note that this value is a local function of $\widetilde{E}^i_a$ evaluated at the spatial point $x$.  Let us now invoke one of the properties of $\psi$ that we have put in place, namely peakedness on a particular configuration $\widetilde{E}^i_a(x)=(\widetilde{E}^i_a(x))_R$ to be determined.  Then evaluated at $(e_{ia})_R$, we have
\begin{eqnarray}
\label{COMMU2}
i(\hbar{G})\Bigl(\frac{\delta{S}}{\delta\widetilde{E}^i_a(x)}\Bigr)_{e_R}=0.
\end{eqnarray}
\noindent
Equation (\ref{COMMU2}) is the condition that the semiclassical momentum is zero for precisely those configurations $e_R$ (namely $e^a_i(x)=(e^a_i(x))_R~\forall{x}\in\Sigma$ or alternatively, the same thing expressed in terms of densitized triads) for which $\psi$ is peaked.  This may seem counterintuitive at first, however it is simply the observation that $e_R$ is, by definition, a critical point of the functional $S$.
\section{5. Initial value constraints}

Let us now evaluate the effect on our Ansatz (\ref{ASSUME}) with respect to the initial value constraints, starting with the kinematic (Gauss' law $G_a(x)$ and spatial diffeomorphism $H_i(x)$) constraints, and then moving on the local Hamiltonian constraint $H(x)$.  The condition that
$\psi\in{Ker}\widehat{H}_i(x),\widehat{G}_a(x)$ is equivalent to the requirement that $\psi$ be gauge and diffeomorphism invariant.  So for arbitrary smooth gauge and diffeomorphism parameters $\theta^a(x),~\xi^i(x)\in{C}^{\infty}(\Sigma)$, we would like to have
\begin{eqnarray}
\label{COMMU3}
\delta_{Kinematic}\psi=\int_{\Sigma}d^3x\Bigl[\frac{\delta\psi[\widetilde{E}]}{\delta\widetilde{E}^i_a(x)}\delta_{\theta}\widetilde{E}^i_a(x)
+\frac{\delta\psi[\widetilde{E}]}{\delta\widetilde{E}^i_a(x)}\delta_{\xi}\widetilde{E}^i_a(x)\Bigr]=0,
\end{eqnarray}
\noindent
where the kinematic transformations act on the densitized triad, respectively via
\begin{eqnarray}
\label{COMMU4}
\delta_{\theta}\widetilde{E}^i_a=-\epsilon_{abc}\widetilde{E}^i_b\theta_c;~~\delta_{\xi}\widetilde{E}^i_a=L_{\xi}\widetilde{E}^i_a=\xi^j\partial_j\widetilde{E}^i_a-(\partial_j\xi^i)\widetilde{E}^j_a,
\end{eqnarray}
\noindent
where $L_{\xi}$ denotes the Lie derivative along the vector field $\xi=\xi^i\partial_i$.  Upon substitution of (\ref{ASSUME}) into (\ref{COMMU3}), and by utilizing the definition of $\psi$, we would need
\begin{eqnarray}
\label{COMMU5}
\delta_{Kinematic}{\psi}=\int_{\Sigma}d^3x\Bigl(\frac{\delta{S}}{\delta\widetilde{E}}^i_a\Bigr)\bigl(\delta_{\theta}\widetilde{E}^i_a+L_{\xi}\widetilde{E}^i_a\bigr)\psi=0.
\end{eqnarray}
\noindent
Let us first analyse the configurations $\widetilde{E}^i_a=(\widetilde{E}^i_a)_R$, upon which $\psi$ is peaked by definition.  Here we have $(\delta{S}/\delta\widetilde{E}^i_a)_{e_R}=0$ due to (\ref{COMMU2}).  Since the kinematic constraints are linear in momenta, the transformations that they generate are sensitive only to the first order of momentum, which from (\ref{COMMU2}) is zero at the Hamilton--Jacobi level since we are now restricted to critical point configurations for $S$.  It is a general result in quantum Hamilton--Jacobi dynamics that there is an inherent semiclassical-quantum correspondence for operators linear in momentum, which becomes broken for operators nonlinear in momenta.  The result is that irrespective of the specific functional form of $\psi$, within the definitions provided, that $\psi$ is exactly gauge-difffeomorphism invariant for $e_{ia}=(e_{ia})_R$.\par
\indent
This leaves remaining those configurations $\widetilde{E}^i_a\neq(\widetilde{E}^i_a)_R$.  Since we are no longer at a critical point of $S$ we can no longer take the kinematic invariances for granted.  The invariance of $\psi$ stemming from (\ref{COMMU5}) requires that $\widehat{G}_a(x)\psi=\widehat{H}_i(x)\psi=0~\forall{x}\in\Sigma$, namely that the Gauss' law and diffeomorphism constraints be satisfied by the state $\psi$.  If these states do not satisfy the kinematic constraints for these configurations, then the invariance is said to be broken.  We will take the conservative approach which subscribes to the notion that by virtue of having a regulator present (which will become necessary for the Hamiltonian constraint), background independence and therefore diffeomorphism invariance could potentially be broken.  However, by definition as an element of $\boldsymbol{S}(\Omega)$ and from its rapid decrease properties, we will be able to ensure that $\psi$ in (\ref{COMMU5}) can be made arbitrarily close to zero.  So in the worst case scenario where the kinematic invariances do indeed become broken, we will be able to ensure that the extent of the violation can be made arbitrarily small by choice of a regulating parameter $\epsilon$ to be specified, which is contained in $S$.  Given the impending prospect of this inherent constraint violation, we must, in due course, provide a prescription whereupon these invariances are ultimately restored within our final results.\par
\indent
Having discussed the situation regarding the kinematic constraints with the impending prospect of constraint violation, we will revisit them later in this paper, and for the time being move on to the Hamiltonian constraint.  Rewriting equation (\ref{DELTA16}) in the functional Schr\"odinger representation, we have
\begin{eqnarray}
\label{COMMU6}
\widehat{H}_{\epsilon}(x)\psi[\widetilde{E}]
=\biggl[e\Bigl(\frac{R}{2}+\Lambda+\frac{9}{2}(\hbar{G}f_{\epsilon}(0))^2\Bigr)+\frac{1}{2}(\hbar{G})^2\widetilde{\epsilon}^{ijk}\epsilon_{abc}
\Bigl(\frac{\delta}{\delta\widetilde{E}^i_a(x)}e^b_j(x)\frac{\delta}{\delta\widetilde{E}^k_c(x)}\Bigr)_{reg}\biggr]\psi[\widetilde{E}]=0,
\end{eqnarray}
\noindent
where we have anticipated the necessity for regularization of the momentum squared terms.  Conceptually speaking, the necessity for regularization is deemed to be a main rationale for the existence of the functional Schwartz space $\boldsymbol{S}(\Omega)$.  Using (\ref{ASSUME}), we have
\begin{eqnarray}
\label{COMMU7}
\widehat{H}_{\epsilon}(x)\psi[\widetilde{E}]
=\biggl[e\Bigl(\frac{R}{2}+\Lambda+\frac{9}{2}(\hbar{G}f_{\epsilon}(0))^2\Bigr)\nonumber\\
+\frac{1}{2}(\hbar{G})^2\widetilde{\epsilon}^{ijk}\epsilon_{abc}
\Bigl[e^a_i\Bigl(\frac{\delta{S}}{\delta\widetilde{E}^j_b}\Bigr)\Bigl(\frac{\delta{S}}{\delta\widetilde{E}^k_c}\Bigr)
+\Bigl(\frac{\delta{e}^b_j}{\delta\widetilde{E}^i_a}\Bigr)_{reg}\Bigl(\frac{\delta{S}}{\delta\widetilde{E}^k_c}\Bigr)
+e^a_i\Bigl(\frac{\delta^2S}{\delta\widetilde{E}^j_b\delta\widetilde{E}^k_c}\Bigr)_{reg}\biggr]\psi[\widetilde{E}]=0.
\end{eqnarray}
\noindent
There are a few observations which can be made regarding (\ref{COMMU7}).  The first term of the second line, in combination with the first line excluding the $(f_{\epsilon}(0))^2$ term, are Hamilton--Jacobi equation-type terms which one would normally attempt to solve for at the semiclassical level.  The remaining terms in the second line will each acquire a factor of $f_{\epsilon}(0)$ due merely to the regularization prescription, and conventionally would be neglected at the semiclassical level.  But what we would prefer ultimately is not just a semiclassical solution but rather, an exact quantum solution to all orders.  Let us catalogue the issues which must be surmounted.\par
\indent
There are two cases which we need to analyse, namely those configurations $e\neq{e}_R$ and $e=e_R$, and we while we would like an exact quantum solution in each case, we will not be able to achieve this on the space $\boldsymbol{S}(\Omega)$.  Moreover, the Hamiltonian constraint poses further challenges pertaining to the  regulator.\par
\indent

\subsection{The Hamiltonian constraint for peaked configurations $e=e_R$}

Let us focus first on the $e=e_R$ case.  In other words, we will assume initially that there is little or nothing known about the wavefunctional $\psi[\widetilde{E}]=e^{S[\widetilde{E}]}$, and first examine the possibility that there exist regions in configuration space $\Omega$ for which  the functional $S[\widetilde{E}]$ has a critical point.\footnote{Note that it is not necessary for the wavefunctional $\psi$ to be peaked for this to be the case.  So the term "peaked configurations" is somewhat of an abuse of terminology.  We will prove that there must indeed be peakeness on $e_R$ as a necessary and sufficient condition to obtain a physical solution $\psi$.  It is the physical solutions which we are interested in in this paper.}  Note, due to (\ref{COMMU2}), that for any $\epsilon>0$ the first and second terms in the second line of (\ref{COMMU7}) will always be zero.  This evaluation must be performed prior to removal of the regulator.  While the first functional derivative of $S$ is zero due to vanishing momentum for $e=e_R$, the second functional derivative must be chosen not to be zero.  So the only momentum squared term that will contribute for these configurations is the third term in the second line of (\ref{COMMU7}).  Let us evaluate this term, inserting the regulator as required
\begin{eqnarray}
\label{COMMU8}
\Bigl(\frac{\delta^2S}{\delta\widetilde{E}^i_a(x)\delta\widetilde{E}^j_b(x)}\Bigr)_{reg}
=\int_{\Sigma}d^3y\tilde{f}_{\epsilon}(x,y)\Bigl(\frac{\delta^2S}{\delta\widetilde{E}^i_a(x)\delta\widetilde{E}^j_b(y)}\Bigr)_{e_R}=\tilde{f}_{\epsilon}(0)M^{ab}_{ij},
\end{eqnarray}
\noindent
where $M^{ab}_{ij}$ is the Hessian matrix of the functional $S$\footnote{This is the second order coefficient in a Taylor expansion of $S$ about the supported configuration $(\widetilde{E}^i_a)_R$.}.\par
\indent
As we have in (\ref{COMMU7}) a Hamiltonian constraint than blows up upon naive removal of the regulator (and certainly a worsening situation as $\epsilon\rightarrow{0}$), it is clear that the only chance of cancelling the singular term and satisfying the constraint is via a sufficiently singular wavefunctional which will bring down the appropriate factor.  Ones hopes that the required $\psi$ is not pathological.  So let us further refine the Ansatz (\ref{ASSUME}), by specifying the required choice of functional $S$.  So we will make the replacement $S\rightarrow\alpha{f}_{\epsilon}(0)S$, hence
\begin{eqnarray}
\label{COMMU9}
\psi_{\epsilon}[\widetilde{E}]=e^{\alpha{f}_{\epsilon}(0)S}
\end{eqnarray}
\noindent
where $\alpha$ is a numerical constant which remains to be determined.  Note that $\psi_{\epsilon}$ becomes labelled by $\epsilon$ and in fact, $f_{\epsilon}(0)$ is the same regularizing functional appearing in (\ref{DELTA7}) and (\ref{DELTA8}) in the coincidence limit.  It may come as no surprise that in order to cancel out regulator dependent terms of the Wheeler--DeWitt equation in this representation, that a regulator-dependent wavefunctional might be needed.  So (\ref{COMMU7}) will have regulator dependence both in the constraint and in $\psi$ and the full solution must be considered in light of removal of the regulator from both quantities.\par
\indent
Performing the replacement of (\ref{COMMU9}) in (\ref{COMMU7}) and continuing with the $e=e_R$ case, then the relevant terms of the regularized Hamiltoinian constraint are
\begin{eqnarray}
\label{COMMU10}
\widehat{H}_{\epsilon}(x)\psi_{\epsilon}[\widetilde{E}]=e\Bigl[\frac{R}{2}+\Lambda+\frac{1}{2}(\hbar{G}f_{\epsilon}(0))^2\bigl(9+\alpha M\bigr)\Bigr]\psi_{\epsilon}[\widetilde{E}],
\end{eqnarray}
\noindent
where we have defined
\begin{eqnarray}
\label{COMMU111}
M=\widetilde{\epsilon}^{ijk}\epsilon^{abc}(e^a_i)_RM_{jbkc}.
\end{eqnarray}
\noindent
We are still required to satisfy the Hamiltonian constraint for the peaked $e=e_R$ configurations.  As $\epsilon\rightarrow{0}$, the regulator term in (\ref{COMMU10}) approaches $(\delta^{(3)}(0))^2$ which is ill-defined.  Note that the choice $\alpha=-9/M$ eliminates the $(f_{\epsilon}(0))^2$ term, $\forall\epsilon$, leaving us with a well-defined condition $R+2\Lambda=0$ which exactly satisfies the Hamiltonian constraint for $all$ $\epsilon>0$.  So we make this choice for $\alpha$ accordingly, with wavefunctional
\begin{eqnarray}
\label{COMMU11}
\psi_{\epsilon}[\widetilde{E}]=\hbox{exp}\Bigl[-\frac{9f_{\epsilon}(0)}MS[\widetilde{E}]\Bigr].
\end{eqnarray}
\noindent
A consistency check confirms that $\alpha$ is negative provided that $M$ is a positive numerical constant.  A positive Hessian matrix confirms that $S$ is peaked on the configurations $e_R$ and that they are stable.  As $f_{\epsilon}(0)$ blows up as $\epsilon\rightarrow{0}$, then for $S>0$ the functional $\psi$ indeed has the fall-off properties desired of $\boldsymbol{S}(\Omega)$.  We have narrowed down the form of the wavefunctional which results in elimination of all regulator dependence in the Hamiltonian constraint.  We would like to remove the regulator by taking the limit $\epsilon\rightarrow{0}$.  But for the time being, the most that can be said is that
\begin{eqnarray}
\label{COMMU12}
\widehat{H}(x)\psi\longrightarrow(R+2\Lambda)\psi_{\epsilon}[\widetilde{E}]=0,~\forall\epsilon>0.
\end{eqnarray}
\noindent
Since the Hamiltonian constraint annihilates the regularized delta functional based on $\alpha=-9/M$, one might be inclined to attempt to evaluate $\widehat{H}(x)\psi=\hbox{lim}_{\epsilon\rightarrow{0}}\widehat{H}_{\epsilon}(x)\psi_{\epsilon}[\widetilde{E}]$ $\epsilon\rightarrow{0}$ in limit of removal of the regulator.  But his limit has not yet been defined.  However, as the choice $\alpha=-9/M$ has resulted in the elimination of any explicit $\epsilon$ dependence in the prefactor multiplying the wavefunctional, which can be interpreted as the spectrum of the Hamiltonian constraint operator, then the requirement that zero be an element of that spectrum is tantamount to the existence of wavefunctionals in $Ker\widehat{H}(x)$.  The result is that the Wheeler-DeWitt equation at the quantum level defined by the affine algebra (\ref{DELTA}) is satisfied by those wavefunctionals with support on triads satisfying $R=-2\Lambda$.  These triads are of constant curvature and include the Schwarzschild--DeSitter solution in isotropic coordinates.  We will show in the next section that the general solution has two degrees of freedom per spatial point $x$.\par
\indent

\subsection{The Hamiltonian constraint for non-peaked configurations $e\neq{e_R}$}

Having evaluated the peaked $e=e_R$ configurations, we are now ready to move on to configurations $e\neq{e}_R$.  At this stage, we have the advantage of knowing a little more regarding the specific form of the wavefunctional $\psi[\widetilde{E}]$ from having considered the previous case.  In particular, $\psi[\widetilde{E}]=\psi_{\epsilon}[\widetilde{E}]$ has picked up some regulator-dependence.  Note, for these $e\neq{e}_R$ configurations, that the quantum terms in the second line of (\ref{COMMU7}) would blow up as $(f_{\epsilon}(0))^2\rightarrow(\delta^{(3)}(0))^2$ if the regulator were to be removed.  However, due to (\ref{TOPO2}), we will be able to arrange it such that as $\epsilon\rightarrow{0}$, $\psi$ falls off faster than the terms in brackets, which are at most polynomial
in $f_{\epsilon}(0)$, will blow up.  This is guaranteed by the form of $\psi$ as determined in (\ref{COMMU9}).  So for $e\neq{e}_R$ the condition $\widehat{H}_{\epsilon}(x)\psi=0$ can be satisfied arbitrarily closely as $\epsilon\rightarrow{0}$.  The limit $\epsilon=0$ would make $\psi$ be the delta functional, which being unnormalizable is no longer contained in $\boldsymbol{S}(\Omega)$.  So even though we will be able to satisfy the Hamiltonian constraint to arbitrary precision for these configurations, we will not be able to achieve convergence to an exact solution on the functional Schwartz space $\boldsymbol{S}(\Omega)$.  We will return to address this in the latter part of the paper.\par
\indent

\subsection{Physical interpretation}
The previous manipulations might suggest that we have chosen a wavefunctional which is separately annihilated by the kinetic and potential terms of the Wheeler--DeWitt equation and is therefore unphysical.\footnote{We have already shown that each term is self-adjoint, and if there each separately annihilate $\psi$ then they would have simultaneous eigenvectors.  But to our knoledge the potential and kinetic operators do not commute.  So it must be the case (as we will ultimately argue) that it is their combined sum that annihilates $\psi$.}  Let us first emphasize that while the momentum $K^a_i$ is zero via (\ref{COMMU2}), its square (schematically, $KeK$) has resulted in a cancellation of the $(f_{\epsilon}(0))^2$ term in (\ref{DELTA16}), and based on that alone clearly is nonzero.  But we would nevertheless like to have more freedom in the choice of the curvature so that it is the clear that it is the sum of the kinetic and potential terms (and not individually) that annihilate the state $\vert\psi\rangle$.  Given that the momentum will always be zero on account of (\ref{COMMU2}) due to the peakedness property of $\boldsymbol{Z}_{e_R}(\Omega)$, then the contribution to $R$ due to the kinetic term must be a quantum effect.  A strong analogy happens for the harmonic oscillator ground state.  The oscillator Hamiltonian $H=\frac{P^2}{2}+\frac{X^2}{2}$ admits a classical solution corresponding to zero energy (the ground state $H=0$) with $x=0,p=0$.  This quantum mechanically is congruous with the expectation values $\langle{X}\rangle=\langle{P}\rangle=0$, in direct analogy to (\ref{COMMU2}).  Quantum mechanically, the quantities $\langle{X}^2\rangle$ and $\langle{P}^2\rangle$ are nonzero which gives rise to a variance, hence the uncertainty principle
\begin{eqnarray}
\label{VARIANCE}
\Delta{X}\Delta{P}=\sqrt{\langle{X^2}\rangle-\langle{X}\rangle^2}\sqrt{\langle{P^2}\rangle-\langle{P}\rangle^2}
\geq\Bigl\vert\frac{1}{2}\langle[X,P]\rangle\Bigr\vert=\frac{\hbar}{2}.
\end{eqnarray}
\noindent
The kinetic term of the oscillator Hamiltonian induces operator ordering ambiguities (in the gravitational case these induce $f_{\epsilon}(0)$ terms), which result in a zero point energy of $\frac{\hbar\omega}{2}$ which is different from the classical value of zero.  The classical solution $x=0,~p=0$ for the oscillator is physically uninteresting and corresponds to a severe reduction of the phase space accessible to the system.  For gravity we have arrived at $R+2\Lambda=0$.  It would be a stretch to regard this as a negligible phase space since, as we will show, it actually contains the two local degrees of freedom per point of GR!  Next, while the oscillator momentum $p$ is zero classically as well as $p^2=0$, the quantum mehanical operator $P^2$ on $\psi(x)$ results in a nonzero zero point energy.  This is a quantum effect which (aside from the cancellation of $(f_{\epsilon}(0))^2$ for peaked configurations $e_R$) appears not to have a quantum gravitational analogue.  But we will demonstrate that one is not limited to $R+2\Lambda=0$, and can actually have solutions with arbitrary constant $R$.\par
\indent
To enforce a generic constant scalar curvature $R=6k$, the result can be easily generalized  by instead taking
\begin{eqnarray}
\label{REPLACE}
\alpha = -\frac{1}{M}\Bigl(\frac{2(3k+\Lambda)}{(\hbar{G}f_{\epsilon}(0))^2}+9\Bigr).
\end{eqnarray}
\noindent
Repeating the previous calculations, everything goes though as detailed except that now, we get an additional contribution to $R$ in (\ref{COMMU10}), wherein $2\Lambda$ gets replaced by $-6k$, and the $(f_{\epsilon}(0))^2$ singularity remains the same as before.  Again, we make the choice $\alpha$ in (\ref{REPLACE}) to cancel the unwanted terms, and we are left with
\begin{eqnarray}
\label{REPLACE1}
\widehat{H}(x)\psi_k\longrightarrow(R-6k)\psi_{k,\epsilon}[\widetilde{E}]=0,
\end{eqnarray}
\noindent
where now the wavefunctional $\psi[\widetilde{E}]$ is given by
\begin{eqnarray}
\label{REPLACE2}
\psi_{k,\epsilon}[\widetilde{E}]=\hbox{exp}\Bigl[-\frac{1}{M}\Bigl(9f_{\epsilon}(0)+\frac{2(3k+\Lambda)}{(\hbar{G})^2f_{\epsilon}(0)}\Bigl)S\Bigr].
\end{eqnarray}
\noindent
Note that as $\epsilon\rightarrow{0}$ that the $k$ contribution becomes negligible and (\ref{REPLACE2}) seems to approach (\ref{COMMU11}).  However, the condition (\ref{REPLACE2}) shows that the peakedness is now with respect to a different curvature, which is a measurable quantum mechanical effect.  Therefore (\ref{REPLACE2}) and (\ref{COMMU11}) are different states, even if one were to remove the regulator.

\section{6. The Yamabe problem}

Amongst the main results of this paper we will characterize the quantum states, for the full theory of gravity, corresponding to manifolds of constant spatial curvature.  This brings to mind an apposite topic, namely the Yamabe problem, which concerns the existence of Riemannian metrics with constant scalar curvature \cite{YAM}.\par
\indent
The Yamabe problem can be stated as follows.  Given a smooth, compact manifold $\Sigma$ of dimension $n\geq{3}$ with a Riemannian metric $g$, there exists a metric $\tilde{g}$ conformal to $g$ for which the scalar curvature $R_{\tilde{g}}$ of $\tilde{g}$ is constant.  As a note of historical interest, the original proof of this theorem due to Yamabe\cite{Yamabe} contained an error, which was found and corrected by Trudinger\cite{TRUD}.  So the theorem stands, and we will apply it to assist in the visualization of our results in the way of concrete solutions to the Einstein equations.  The Yamabe problem is given by the equation governing conformally related curvature scalars
\begin{eqnarray}
\label{YAMABE}
-4\Bigl(\frac{n-1}{n-2}\Bigr)\Delta_g\varphi+R_g\varphi=R_{\tilde{g}}\varphi^{\frac{n+2}{n-2}},
\end{eqnarray}
\noindent
where $\tilde{g}=\varphi^{\frac{4}{n-2}}g$ is the conformally related metric with $\varphi\in{C}^{\infty}(\Sigma),~\varphi>0$.  $\Delta_g$ is the Laplacian operator with respect to the metric $g$.  For the purposes of this paper we will take $n=3$ for three dimensional spatial manifolds, with the condition that $R_{\tilde{g}}=6k$ is a constant, hence (\ref{YAMABE}) simlifies to
\begin{eqnarray}
\label{YAMABE1}
\Delta_g\varphi -\frac{R_g}{8}\varphi=-\frac{3k}{4}\varphi^5.
\end{eqnarray}
\noindent
This has solution for $\varphi$ and serves to fix it in the conformal metric $\tilde{g}=\varphi^4g$. Based upon the natural canonical separation of the 3-metric into an intrinsic time $T=\hbox{ln} q^{1/3}$ and unimodular metric ${\bar q}_{ij}=q^{-1/3}q_{ij}$   degrees of freedom  (such that $\hbox{det}({\bar q}_{ij})=1$), classical and quantum gravity with a physical Hamiltonian have been formulated in Ref.\cite{Soo-Yu}. Making the identifications $g\sim{\bar q}_{ij}$ and $\tilde{g}\sim{q}_{ij}$ in (\ref{YAMABE}), the Yamabe problem for a given topology guarantees the existence of solutions for generic ${\bar q}_{ij}$.  The physical degrees of freedom of the graviton are thus contained in the unimodular part of the  spatial metric ${\bar q}_{ij}$ and hence in the unimodular part of the vierbein $e^a_i$ .  Hence this information is encapsulated in the solution of the Yamabe problem  \footnote{For the purposes of normalization, there is no integration with respect to $T(x)=\hbox{ln} q(x)^{1/3}$ which plays the role of a time variable on configuration space, because in quantum mechanics one does not normalize a wavefunction with respect to time.  So the integration measure for $\psi[\widetilde{E}]$ should be taken with respect to the unimodular degrees of freedom of the triad.}.\par
\indent
A general manifold of constant 3D scalar curvature admits two degrees of freedom per spatial point, which can be seen from the point of view of the Yamabe problem, or perhaps more explicitly, from its linearization.  Performing a linearization about a flat Euclidean metric $g_{ij}=\delta_{ij}+h_{ij}$ for example, the 3D spatial scalar curvature $R=\partial_i\partial_jh_{ij}-\partial^2h$, where $h=\delta_{ij}h_{ij}$ is invariant under infinitesimal general coordinate transformations $\delta{h}_{ij}=\partial_i\xi_j+\partial_j\xi_i$.  Starting from a general perturbation $h_{ij}$ with six degrees of freedom and using the coordinate freedom to fix a coordinate system yields $6-3=3$ degrees of freedom.  Then the condition of constant curvature reduces them to $3-1=2$  (transverse traceless) degrees of freedom per point.  So the states $\psi_{k,\epsilon}[\widetilde{E}]$ do indeed capture two local degrees of freedom per point in GR, thus encompassing a wide range of the interesting solutions.  So as alluded to earlier, our quantization procedure indeed captures the essential degrees of freedom of four dimensional gravity.

\par
\indent

\section{7. The Gaussian case}

We have so far analysed the action of the initial value constraints in the functional Schwartz space $\boldsymbol{S}(\Omega)$, finding that the constraints are satisfied "on the average".  More precisely, for triad configurations which peak on constant curvature $R=6k$, the Hamiltonian constraint is identically satisfied for certain $\psi_{k,\epsilon}[\widetilde{E}]\in\boldsymbol{S}(\Omega)$.  While all explicit regulator dependence $f_{\epsilon}(0)$ has disappeared from the Wheeler--DeWitt equation, $\forall\epsilon$, for these states, that dependence resides explicitly within the state.  For $e\neq{e}_R$, we argued that while the $(f_{\epsilon}(0))^2$ terms are no longer cancelled out from $\widehat{H}(x)\vert\psi\rangle=0$, they are suppressed as $\epsilon\rightarrow{0}$ due to the fall-off property (\ref{COMMU2}) of $\boldsymbol{S}(\Omega)$.\par
\indent
To make this concrete, let us use a normalized Gaussian centered on $e=e_R$ as an explicit example
\begin{eqnarray}
\label{GAUSSIAN}
\psi_{\epsilon,k}[\widetilde{E}]
=\Bigl(\prod_{x\in\Sigma}\frac{|\alpha|{f}_{\epsilon}(0)Det{M_R}}{\pi}\Bigr)^{1/4}
\hbox{exp}\Bigl[\frac{\alpha{f}_{\epsilon}(0)}{2}\int_{\Sigma}d^3x
\bigl(\widetilde{E}^i_a(x)-(\widetilde{E}^i_a(x))_R\bigr){M}^{ab}_{ij}
\bigl(\widetilde{E}^j_b(x)-(\widetilde{E}^j_b(x))_R\bigr)\Bigr],
\end{eqnarray}
wherein $Det{M_R}$ denotes $Det(M_{iajb})$ evaluated at $e_{ia} =e^R_{ia}$. To guarantee positivity of the eigenvalues of the Hessian matrix in the Gaussian and to eliminate local internal gauge degrees of freedom, we express the Gaussian wavefunction in the form
\begin{eqnarray}
\label{GAUSSIAN2}
\psi_{\epsilon,k}[\widetilde{E}]
=\Bigl(\prod_{x\in\Sigma}\frac{|\alpha|{f}_{\epsilon}(0)Det{M_R}}{\pi}\Bigr)^{1/4}
\hbox{exp}\Bigl[\frac{\alpha{f}_{\epsilon}(0)}{2}\int_{\Sigma}d^3x
\bigl(\widetilde{q}^{ij}(x)-\widetilde{q}^{ij}_R(x)\bigr){M}_{ijkl}
\bigl(\widetilde{q}^{kl}(x)-\widetilde{q}^{kl}_R(x)\bigr)\Bigr],
\end{eqnarray}
\noindent
with doubly densitized  $\widetilde{q}^{kl} := \widetilde{E}^{ia}\widetilde{E}^{j}_{a}(x)$. The dependence of the wavefunction on $\widetilde{E}^{ia}$ only through the spatial metric ensures the state is explicitly invariant under $SO(3)$ gauge transformations.
The Gaussian can be seen as the terms of a general $S$ up to quadratic order in a Taylor expansion about $e_{ia}$ and $e^R_{ia}$, or rather about $q_{ij}= (q^R_{ij})$ which defines $(e_{ia}, e^R_{ia})$ up to arbitrary $SO(3)$ rotations.  For illustrative purposes, we provide one particular example of a Hessian
\begin{eqnarray}
\label{GAUSSIAN3}
 M_{ijkl}=(1/e^3) [\frac{1}{2}(q_{ik}q_{jl} +  q_{il}q_{jk}) - \frac{\lambda}{3\lambda -1} q_{ij}q_{kl}],
\end{eqnarray}
wherein $M_{ijkl}$ is a generalized DeWitt supermetric with deformation parameter $\lambda$ and signature $[\frac{1}{3} -\lambda, +,+, + ,+,+]$  which can be made positive-definite with the choice $\lambda <\frac{1}{3}$.
The identity
\begin{eqnarray}
\label{Derivative}
\frac{\delta\psi}{\delta\widetilde{E}^i_a(x)}=\int_\Sigma d^3y \frac{\delta\psi}{\delta\widetilde{q}^{jk}(y)}\frac{\delta\widetilde{q}^{jk}(y)}{\delta\widetilde{E}^i_a(x)} = 2\widetilde{E}^{ja}(x) \frac{\delta\psi}{\delta\widetilde{q}^{ij}(x)}
\end{eqnarray}
ensures that (\ref{COMMU2}) holds at the maxima of $\psi$ at constant scalar curvature configurations $q_{ij}= (q^R_{ij})$. The invertible relation between $M_{iajb}$ and $M_{ijkl}$  is  $M_{iajb} = 4M_{kilj}\widetilde{E}^k_a\widetilde{E}^l_b=
\frac{4}{e}[\frac{1}{2}(\delta_{ab}q_{ij} +e_{ja}e_{ib}) - \frac{\lambda}{3\lambda -1}e_{ia}e_{jb}]$, which according to (\ref{COMMU111}), yields  $M >0$ for $\frac{1}{5} <\lambda <\frac{1}{3}$.\par
\indent
The Gaussian (\ref{GAUSSIAN}) explicitly breaks diffeomorphism invariance.  There is no way to make the exponent a density of weight one, since the configuration $(\widetilde{E}^i_a(x))_R$ is inert with respect to functional variations while $\widetilde{E}^i_a(x)$ is not. That notwithstanding, (\ref{GAUSSIAN}) respects the condition (\ref{TOPO2}), and it should be now phsically and mathematically clear the rationale for the desired fall-off properties.  It is because they are tied directly to the regulator $f_{\epsilon}(0)$, whose inverse (in the Gaussian case) is directly related to the width about $e=e_R$.  As $\epsilon\rightarrow{0}$ this width becomes arbitrarily small and while the $e\neq{e}_R$ case is not an exact solution to the constraints, it is an extremely good approximation which gets better and better as $\epsilon\rightarrow{0}$, and $e\neq{e}_R$ configurations become rapidly suppressed. It is also to be noted that as $\epsilon$ and the width of the Gaussian decrease to zero, $(\widetilde{q}^{ij}-\widetilde{q}^{ij}_R)$  tends to the cotangent element $\delta\widetilde{q}^{ij}(x)$ at $q^R_{ij}$, and the wavefunction becomes just the exponentiation (modulo a multiplicative constant) of the explicitly diffeomorphism and gauge invariant ``superspace interval" $ds^2_{\rm Superspace}=\int_{\Sigma}d^3x\,  {M}_{ijkl}(x)\delta\widetilde{q}^{ij}(x)\delta\widetilde{q}^{kl}(x)$ evaluated at $q^R_{ij}$. Similarly, in the same limit of zero width, $(\widetilde{E}^i_a-(\widetilde{E}^i_a)_R)$ tends to the cotangent element $\delta\widetilde{E}^i_a(x)$, and expression (\ref{GAUSSIAN2}) coincides with  (\ref{GAUSSIAN}) with the aforementioned relation between $M_{iajb}$ and $M_{ijkl}$ \footnote{Generalization to non-vanishing momentum values can be made by adding a phase $e^{\frac{i}{\hbar}\int_{\Sigma} (P^R_{ij}(x)\tilde q^{ij}(x))d^3x }$ to the Gaussian of (\ref{GAUSSIAN2}), wherein $ P^R_{ij}(x)$  is restricted by $\int_{\Sigma} (P ^R_{ij}(x)L_{\xi}\tilde q^{ij}_R(x)) d^3x =0 \,\, \forall\vec{\xi}$ , so that spatial diffeomorphism invariance is regained in the limit vanishing $\epsilon$.}.\par
\indent
We have been careful so far in this paper not to use the expression $\hbox{lim}_{\epsilon\rightarrow{0}}$, even though we can make $\epsilon$ arbitrarily close to zero.  Note that we have included a normalization pre-factor in (\ref{GAUSSIAN}), upon which once can base a Hilbert space structure.  This is what makes the Gaussian special, the fact that they are well-known as minimum uncertainty states.  Let us use the following shorthand notation to denote the Gaussian state (\ref{GAUSSIAN})
\begin{eqnarray}
\label{DENOTE}
\vert{k}\rangle_{\epsilon}\sim{e}^{-\alpha_k(\widetilde{E}-\widetilde{E}_{R(k)})^2}
\end{eqnarray}
\noindent
with no loss of generality at it applies to infinite dimensional spaces.  It is taken to be understood in (\ref{DENOTE}) that the width of the Gaussian is tied directly to the regulating function $f_{\epsilon}(0)$ as well as the curvature constant $k$.  One can construct an overcomplete basis of coherent states peaked on constant curvature solutions for each $\epsilon>0$.\footnote{This brings to mind, for example, the coherent states peaked on particular geometries, which Loop Quantum Gravity uses to attempt to investigate the semiclassial limit of spin networks.  In the functional Schwartz space $\boldsymbol{S}(\Omega)$, the semiclassical limit would be abundantly clear for the reasons we have explained.}
\begin{eqnarray}
\label{DENOTE1}
\langle{k}_q\vert{k}_2\rangle_{\epsilon}\sim\sqrt{\frac{1}{w}}e^{-w(\widetilde{E}_{k_1}-\widetilde{E}_{k_2})^2},
\end{eqnarray}
\noindent
where we have defined the inverse of the composite width, which is bounded from below by $sup\vert{k}_1,k_2\vert$
\begin{eqnarray}
\label{DENOTE2}
w=\frac{\alpha_{k_1}\alpha_{k_2}}{\alpha_{k_1}+\alpha_{k_2}}.
\end{eqnarray}
\noindent
Note the Gaussian decay with respect to distance between the triads.  There will always be some nonzero overlap between Gaussian states,  which gets closer to zero the farther apart their curvatures become.  But for $\epsilon=0$ the Gaussian turns into a delta functional
\begin{eqnarray}
\label{DELTAF}
\hbox{lim}_{\epsilon\rightarrow{0}}\psi_{\epsilon,k}[\widetilde{E}]
=\prod_{a,i,x}\delta\bigl(\widetilde{E}^i_a(x)-(\widetilde{E}^i_a)_{R(k)}\bigr)\equiv\delta(\tilde{E}-\tilde{E}_{R(k)}).
\end{eqnarray}
\noindent
As the delta functional is not normalizable, it is not an element of the functional Schwartz space $\boldsymbol{S}(\Omega)$ and neither is it an element of the Hilbert space $\boldsymbol{H}_{Kin}$ in which it is densely embedded.  This means that the limit $\epsilon=0$ doesn't actually exist on $\boldsymbol{S}(\Omega)$, and for this reason we are stuck with having the regulator present while operating in this space, in combination with its associated implications\footnote{Including the possible violation of full gauge-diffeomorphism invariance, however small that violation may be.} which will be tackled in the next section.\par
\indent

\subsection{The rigged, physical Hilbert space}
The condition $\epsilon=0$ is precisely the limit which corresponds to removal of the regulator which, so far, has not been defined.  We have discussed only two components of the rigged Hilbert space $\phi\subset\boldsymbol{H}_{Kin}\subset\phi'$.  We are now ready to bring in the third component $\phi'$, the topological dual of $\phi$ constituting the extension of $\boldsymbol{H}_{Kin}$ to include the so-called ``generalized" functionals.  We now invoke the properties of rigged Hilbert spaces which were established to place these cases on rigorous mathematical footing \cite{VON}, \cite{VON1}, \cite{VON2} by declaring that it is in $\phi'$ where the exact solutions to all of the constraints, namely the $\epsilon={0}$ limit, reside.
\par
\indent
Since the Gaussians (\ref{DENOTE}) smoothly approach a delta functional, then their labels get carried over into the space $\phi'$, where they remain intact.
  We can immediately construct the following states as solutions to all of the constraints
\begin{eqnarray}
\label{VONAGE}
\vert\widetilde{E}_R\rangle=\prod_{a,i}\prod_{x\in\sigma}\vert(\widetilde{E}^i_a(x))_R\rangle
\end{eqnarray}
\noindent
with the inner product
\begin{eqnarray}
\label{VONAGE1}
\langle\widetilde{E}_{R(k_1)}\vert\widetilde{E}_{R(k_2)}\rangle=\delta(E_{R(k_1)}-E_{R(k_2)})
=\prod_{a,i}\prod_{x\in\Sigma}\delta\bigl((\widetilde{E}^i_a(x))_{R(k_1)}-(\widetilde{E}^i_a(x))_{R(k_2)}\bigr).
\end{eqnarray}
\noindent
Note that (\ref{VONAGE}) and (\ref{VONAGE1}) are inert with respect to functional variations.  Therefore they represent gauge-diffeomorphism invariant objects.  These variances, which may or may not have been broken on $\phi$, are by definition automatically restored on the space $\phi'$, where the wavefunctionals are infinite spikes with support on the constant curvature configurations.  The statement of constant curvature $R=6k$ is gauge-diffeomorphism invariant, as the Lie derivative $L_{\xi}R=\xi^i\partial_iR=0$ when $R$ is constant.\par
\indent
Such delta-functional normalizable states are naturally occurring in quantum mechanics for self adjoint unbounded operators with a continuous spectrum (in this case, the value of $k$).  The states are continuous rather than Kronecker delta normalizable, since they represent two local decrees of freedom per point as labelled by the Einstein solutions $e_R$.  Delta functional normalizability is sufficient for our purposes since normalized physical states can be constructed from superpositions of them.  So the constant curvature manifold states constitute a basis for this expansion.  These states are also annihilated exactly by the Hamiltonian constraint, hence providing a possible solution to the decades-old problem of making sense of the Wheeler--DeWitt equation \cite{DeWitt-Wheeler}.\par
\indent
Given that the aforementioned states solve the constraints, a natural question is whether there exists a gauge invariant, diffeomorphism invariant measure on the space of densitized triads.  One may be inclined to use the naive infinite dimensional Lebesgue measure in the formal sense in manipulating the bras and kets.  However, explicitly proving these invariances is a nontrivial task, which is required in order to for the invariances to apply to the states.\par
\indent
We provide a proposition for a measure which implements these invariances unitarily, which we relegate to the next article in this series.  The proposition is to use the Lebesgue measure of the unimodular part of the inverse triad $\overline{E}^i_a=(\hbox{det}e)^{1/3}E^i_a$ containing the two local physical degrees of freedom per spatial point (here $E^i_ae^a_j=\delta^i_j$ and $E^i_ae^b_i=\delta^b_a$).  The unimodular form of the associated deWitt metric on this configuration space $\Omega$ has determinant one and is therefore gauge-diffeomorphism invariant.  Finally, we will apply the terminology ``Rigged Physical Hilbert Space" to this set of delta-functional normalizable states with support on constant curvature manifolds.  Besides the fact that there states capture the two local degrees of freedom per spatial point and admit an appropriate Hilbert space structure, they as well capture a variety of interesting solutions to the Einstein equations.\footnote{For example, the blackhole, amongst other solutions, have constant curvature foliations.}

\section{Remarks}

The results of this work show that challenging task of finding exact solutions in the full theory quantum gravity can find resolution. 
We have obtained exact solutions of the Wheeler-DeWitt equation and have constructed a rigged physical Hilbert space $\phi'$ which can be tied directly to the Yamabe construction.
Thus these states explicitly capture the two essential local degrees of freedom per point of GR, and are gauge-diffeomorphism invariant. Moreover, we have elucidated upon the interpretation of regulator-dependence on the Hilbert space version of the states, and how this manifests itself with respect to spatial geometries. It remains as a future research direction to analyze the superspace geometry on $\phi'$ with respect to the physical degrees of freedom, taking into account the measure on the space of states\footnote{The proposition is to base the superspace dynamics upon the unimodular degrees of freedom of the triads and analogously for the superspace metric.}.

In quantum field theory, Newton-Wigner states\cite{Newton-Wigner} form an important bridge between the description of matter with quantum fields at the fundamental level, and the localized particle description of classical physics. These Newton-Wigner wavefunctions  are ``localized" in that any state which is infinitesimally displaced in spatial coordinate becomes orthogonal to itself i.e. two states which differ by infinitesimal spatial displacement have ``no overlap"; and there is one-to-one correspondence between the labels of the states and coordinates of the spatial manifold. Our solutions exhibit some parallel features: these mutually orthogonal states are each parametrized by a different $q^R_{ij}$ which is also a configuration datum of the allowed classical initial phase space data; and because of the local degrees of freedom, two different $q^R_{ij}$ configurations can differ by (a possibly infinitesimal) transverse traceless $h^{TT}_{ij}$ perturbation.
Linearity of the Wheeler-DeWitt operator ensures that arbitrary superpositions of these solutions will also satisfy the quantum constraints. 

It is also noteworthy that
our wavefunctions solve the Wheeler-DeWitt equation in the limit of vanishing $\epsilon$ and in this same limit the diffeomorphism constraint is also satisfied. In the Dirac algebra a super-momentum operator results from the commutator of two super-Hamiltonian operators; thus if the Dirac algebra carries over to the quantum context in the absence of anomalies, a solution of the Wheeler-DeWitt equation must consistently also automatically solve the diffeomorphism constraint. This feature is indeed exhibited by our solutions.

\section{Acknowledgements}
The authors are grateful to Chou Ching--Yi for her analysis, checking and contributions
to some of the various versions of the draft during certain stages of its development.
She has made some notable contributions leading to the final form of the document.
This work has been supported in part by the
Office of Naval Research under Grant No. N-000-1414-WX-20789, and
in part by Perimeter Institute for Theoretical Physics (research at Perimeter
Institute is supported by the Government of Canada through Industry
Canada and by the Province of Ontario through the Ministry of Economic
Development and Innovation); and also supported in part by the National Science Council of Taiwan under Grant No.
NSC101-2112-M-006 -007-MY3, and the National Center for Theoretical
Sciences, Taiwan.  And finally but not least of all, the authors would like to thank Lee Smolin
for doing a critical read of the manuscript, along with his helpful suggestions and insights.

\end{document}